\documentstyle[aps,epsf,twocolumn]{revtex}

\def\be{\begin{equation}}
\def\ee{\end{equation}}
\def\bea{\begin{eqnarray}}
\def\eea{\end{eqnarray}}

\def\rd{\mbox{d}}

\def\re{\mbox{e}}
\def\t12h{\frac{\theta_{12}}{2}}

\def\eps{\varepsilon}

\def\r#1{(\ref{#1})}
\def\nn{\nonumber\\}

\def\NPB#1#2#3{{ Nucl. Phys.} {\bf B#1} (#2) #3}
\def\PRD#1#2#3{{ Phys. Rev.} {\bf D#1} (#2) #3}
\def\PRB#1#2#3{{ Phys. Rev.} {\bf B#1} (#2) #3}
\def\PRL#1#2#3{{ Phys. Rev. Lett.} {\bf #1} (#2) #3}
\def\TMP#1#2#3{{ Theor. Mat. Phys.} {\bf #1} (#2) #3}

\begin{document}

\preprint{OUTP-97-23S}

\title{\bf Dynamical Magnetic Susceptibilities in Cu Benzoate}
\author {Fabian H.L. Essler and Alexei M. Tsvelik}
\address{Department of Physics, Theoretical Physics,
        Oxford University\\ 1 Keble Road, Oxford OX1 3NP, United Kingdom}

\maketitle
\begin{abstract}
Recent experiments on the quasi 1-D antiferromagnet Cu Benzoate
revealed a magentic field induced gap coexisting with (ferro)magnetic
order. A theory explaining these findings has been proposed by
Oshikawa and Affleck. In the present work we discuss
consequences of this theory for inelastic neutron scattering
experiments by calculating the dynamical magnetic susceptibilities
close to the antiferromagnetic wave vector by the formfactor method.

\end{abstract}

\begin{narrowtext}
\section{Introduction}
It has been known for some time that Cu Benzoate is a quasi-1D $S=1/2$
(Heisenberg) antiferromagnet \cite{date}. In a recent neutron
scattering experiment \cite{dender} in a magnetic field,
the existence of field-dependent incommensurate low energy modes was
established. The incommensurability was found to be consistent with
the one predicted by the exact solution of the Heisenberg model in a
magnetic field. However, the system exhibited an unexpected excitation
gap induced by the applied field. These findings were quite surprising
as the coexistence of a gap and a magnetization is inconsistent with
rotational symmetry around the direction of the magnetization
\cite{oa2}.

The resolution of this puzzle goes as follows: Dender {\sl et al}
\cite{dender} suggested that the gap could be due to the staggered
magnetic field generated by the alternating $g$-tensor in Cu Benzoate; 
Oshikawa and Affleck \cite{oa} then showed that this mechanism (in
particular the generation of a staggerd field {\sl perpendicular} to
the direction of the applied magnetic field), together with the
Dzyaloshinskii-Moriya interaction present in Cu Benzoate, indeed can
account for the experimental findings. They proposed that Cu Benzoate
is described by the Hamiltonian 
\be
H=\sum_i J \vec{S}_i\cdot\vec{S}_{i+1}- H S_i^z + h (-1)^i S^x_i\ ,
\label{hamil}
\ee
where $H\gg h$. The low energy effective theory of \r{hamil} is
obtained by abelian bosonization and is given by a Sine-Gordon model
with Lagrangian density \cite{oa}
\be
{\cal L} = \frac{1}{2} (\partial_\mu\Phi)^2 + \lambda \cos(\beta
\ \Theta)\ .
\label{lagr}
\ee
Here $\Theta$ is the dual field and fulfils
$\partial_x\Phi = -\partial_t\Theta$, $\partial_t\Phi =
-\partial_x\Theta$. This implies that $(\partial_\mu\Phi)^2
=(\partial_\mu\Theta)^2 $ and \r{lagr} thus is a Sine-Gordon model for
the dual field. We note that all of our analysis is based on the model
\r{hamil} and \r{lagr} so that the results carry over to any material
described by this theory (with appropriate fit of $\frac{H}{J}$ to
experiment).

The coupling constant $\beta$ is a function of the applied field
$H$ and is related to the ``dressed charge'' (see \cite{vladb} and
references therein) of the isotropic Heisenberg chain in a magnetic
field by
\be
\beta= \frac{\sqrt{\pi}}{Z}\ .
\ee
For further convenience we define the quantity
\be
\xi=\frac{\beta^2}{8\pi-\beta^2}\ .
\label{xi}
\ee
The dressed charge can be determined exactly from the Bethe Ansatz
solution. It is given by $Z=Z(A)$, where $Z(\lambda)$ is a solution of
the integral equation \cite{vladb}
\be
Z(\lambda)+\frac{1}{2\pi}\int_{-A}^A d\mu\
\frac{2}{1+(\lambda-\mu)^2}\ Z(\mu) = 1\ .
\ee 
The integration boundary $A$ is a function of the magnetic field $H$
and is determined by the condition $\eps(A)=0$, where
\be
\eps(\lambda)+\frac{1}{2\pi}\int_{-A}^A d\mu\
\frac{2}{1+(\lambda-\mu)^2}\ \eps(\mu) =
\frac{2H}{J}-\frac{2}{\lambda^2+\frac{1}{4}}\ . 
\ee 
These integral equations can easily be solved numerically. In
Fig.~\ref{fig:beta} we plot $\beta^2/2\pi$ of (where $\beta$ is the
coupling constant of the Sine-Gordon model \r{lagr}) as a function of
the applied magnetic field $H$ for Cu Benzoate. We note that at zero
field $\beta^2/2\pi=1$.
\begin{figure}[ht]
\begin{center}
\noindent
\epsfxsize=0.45\textwidth
\epsfbox{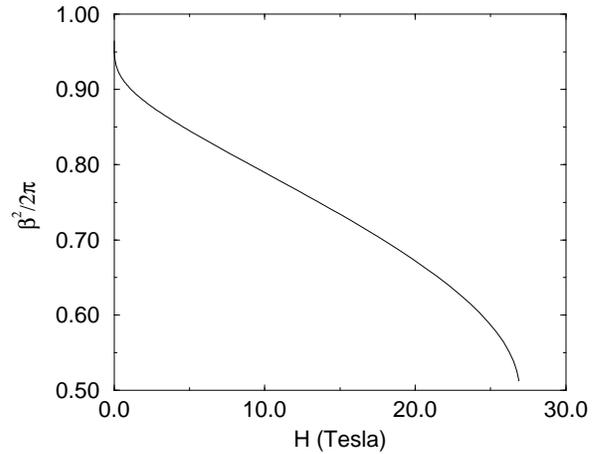}
\end{center}
\caption{\label{fig:beta}%
Coupling constant $\beta^2/2\pi$ as a function of the applied magnetic
field.} 
\end{figure}
The curve in Fig.~\ref{fig:beta} terminates at a value of $H\approx
27$ Tesla ($H=2J$) at which the underlying Heisenberg model
experiences a transition to the saturated ferromagnetic state.

The spectrum of the Sine-Gordon theory \r{lagr} is well known
\cite{spectrum} (see also \cite{zamo2,mtm}): it consists of a
soliton-antisoliton doublet of mass $M$ and their bound states which
are called ``breathers''.
For Cu Benzoate the soliton mass as a function of the applied magnetic
field was determined by Oshikawa and Affleck \cite{oa}
\be
M \approx 1.85 (\frac{h}{J})^{1/(2-\beta^2/4\pi)}J\ ,
\label{solitonmass}
\ee
where the induced staggered field $h$ is determined as a function of
the applied field $H$ (and the Dzyaloshinskii-Moriya interaction in Cu
Benzoate) in \cite{oa}. 

In addition there are $[\frac{1}{\xi}]$ (here $[x]$ denotes the
largest integer smaller than $x$) breathers, denoted by
$B_1,B_2,\ldots$, with masses 
\be
M_n = 2M\ \sin n\pi\xi/2\ ;\  n=1,\ldots ,\left[1/\xi\right]. 
\label{mass}
\ee
The mass spectrum as a function of magnetic field for CuBenzoate is
plotted in Fig.~\ref{fig:breathermass}
\begin{figure}[ht]
\begin{center}
\noindent
\epsfxsize=0.45\textwidth
\epsfbox{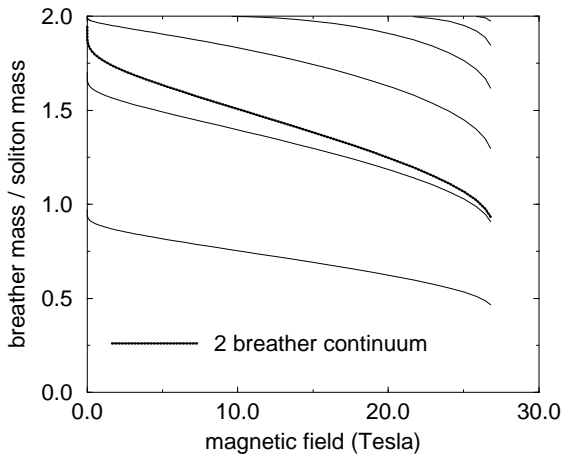}
\end{center}
\caption{\label{fig:breathermass}%
Breather masses in units of the soliton mass as functions of the
applied magnetic field.} 
\end{figure}
We see that for small fields there are {\sl three} breathers, one of
which is just below the soliton-antisoliton continuum starting at
$2M$. For higher fields further breathers split off this
continuum. However they appear at energies higher than the threshold
of the $B_1B_1$ two breather continuum. For small fields the
dependence of the mass ratios $\frac{M_n}{M}$ on $H$ is
\be
M_n\sim 2M\left[ \sin n\pi/6 +\frac{n\pi}{9\ln h/Jh_0} \cos n\pi/6
\right],
\ee
where $h_0=\sqrt{8\pi^3/e}$.

The Lagrangian (\ref{lagr}) can be brought to standard Sine-Gordon
form by performing the duality transformation $\Theta\to\Phi$. The
bosonized expressions of the spin operators change under this duality
transformation as well. We find \cite{aff,book,nst} $\vec{S}_n
\longrightarrow a_0 \vec{S}(x), \ x=na_0$,

\bea
\vec{\bf S} (x) &=& \vec{\bf J}(x) + (-1)^n \vec{\bf n} (x)\ ,\nn 
J^z&=&\frac{1}{\beta}\partial_x\Theta\ ,\nn 
J^+&=&\frac{1}{2\pi a_0} \exp\left( -i\beta\Phi\right)\
\cos\left(\frac{2\pi}{\beta}\Theta-2\delta x\right),\nn 
n^x (x) &=& \frac{\Lambda}{\pi a_0} \cos(\beta\Phi (x)),\nn
n^y (x) &=& \frac{\Lambda}{\pi a_0} \sin(\beta\Phi (x)),\nn
n^z (x) &=& \frac{\Lambda}{\pi a_0} \cos(\frac{2\pi}{\beta}\Theta
(x)-2\delta x).
\label{boso}
\eea
Here $n^a(x)$ are the components of the staggered magnetization,
$J^{\pm,z}$ are the current operators, $na_0=x$ and $\Lambda$ is a
nonuniversal coefficient. The quantity $\delta$ is the difference
between the Fermi momenta for up spins and down spins. It is easily
determined from the exact solution of the XXX Heisenberg chain in a
magnetic field (see \cite{vladb} and references therein) but is not
important for the present work.

The dynamical magnetic susceptibilities for wavevectors close to the
antiferromagnetic wave vector $\pi$ ($\pi\pm 2\delta$ for the
longitudinal part $\chi^{zz}$) are then given by
\bea
\chi^{xx}(\omega,q)&\propto&\int_{-\infty}^\infty dx\int_0^\infty dt
\ e^{i (\omega+i\epsilon)t-i(q-\pi)x}\ \times\nn 
&&\times \langle [\cos\beta\Phi(t,x) ,
\cos\beta\Phi(0,0)]\rangle \nn 
\chi^{yy}(\omega,q)&\propto&\int_{-\infty}^\infty dx\int_0^\infty dt
\ e^{i (\omega+i\epsilon)t-i(q-\pi)x}\ \times\nn 
&&\times \langle[\sin\beta\Phi(t,x) ,
\sin\beta\Phi(0,0)]\rangle \nn 
\chi^{zz}(\omega,q)&\propto&\int_{-\infty}^\infty dx\int_0^\infty dt 
\ e^{i (\omega+i\epsilon)t-i(q-\pi)x}\ \times\nn 
&&\hskip-20pt\times \langle[\cos(\frac{2\pi}{\beta}\Theta(t,x)-2\delta x) ,
\cos\frac{2\pi}{\beta}\Theta(0,0)]\rangle .
\label{suscept}
\eea

The transverse susceptibilities $\chi^{xx}$ and $\chi^{yy}$ can be
determined exactly through the formfactor approach to quantum
correlation functions \cite{karowski,smirnov,lukyanov} (our discussion
follows \cite{etd}). The longitudinal susceptibility involves the dual
field which is a nonlocal operator and cannot be treated in the same
way. We therefore restrict our analysis to the transverse
susceptibilities. As the longitudinal modes become soft at the
incommensurate wavevectors $\pi\pm 2\delta$ rather than at the
Neel wave vector $\pi$ (where the transverse modes are soft) it is
possible to separate $\chi^{zz}$ from $\chi^{\alpha\alpha}\ ,\
\alpha=x,y$ experimentally by performing constant wavevector scans
around $q=\pi\pm 2\delta$ and $q=\pi$ respectively.

Inserting a resolution of the identity in the
correlation functions \r{suscept} we obtain formfactor expansions of
the form 
\bea
\chi^{xx}(\omega,q)&=& -2\pi\sum_{n=1}^\infty\sum_{\eps_i}\!\int\!
\frac{\rd\theta_1\ldots\rd\theta_n}{(2\pi)^nn!}
|F^{cos}_{\eps_1\ldots\eps_n}(\theta_1\ldots\theta_n)|^2\nn
&&\times\bigg\lbrace\frac{\delta[(\pi-q)-\sum_jM_j\sinh\theta_j]}{\omega -
\sum_j M_j\cosh\theta_j +i\epsilon}\nn
&&\qquad -\frac{\delta[(\pi-q)+\sum_jM_j\sinh\theta_j]}{\omega + \sum_j
M_j\cosh\theta_j +i\epsilon}\bigg\rbrace ,\nn
\label{dcos}
\eea
and an analogous expansion for $\chi^{yy}$. The quantities $F^{cos}$
in \r{dcos} are the formfactors of the operator $\cos\beta\Phi$ and
have been calculated exactly \cite{karowski,smirnov,lukyanov}.
The expansion \r{dcos} is in terms of multiparticle states. In order
to make contact with the neutron scattering experiments we are
interested in the imaginary parts of the dynamical susceptibilities,
which are proportional to the neutron scattering cross
sections. Because all elementary excitations in the Sine Gordon theory
have a gap, $n$-particle states will contribute to $\Im
m\chi^{\alpha\alpha}(\omega,q)$ ($\alpha=x,y$) only at energies larger
than the $n$-particle gap. In other words, by taking into account only
the first few terms in the expansion \r{dcos} we obtain an exact
expression for $\Im m\chi^{\alpha\alpha}(\omega,q)$ for energies below
a threshold that depends on the omitted terms. 

As the Sine-Gordon theory is Lorentz invariant the final answer for
the susceptibilities cannot depend on $\omega$ and $q$ independently
but only on the combination
\be
s=\sqrt{\omega^2-(\pi-q)^2}\ .
\ee

The formfactor expansion is now readily performed using the results of
\cite{karowski,smirnov,lukyanov} and the bootstrap procedure to
determine breather formfactors. Using that $\cos\beta\Phi$ is even
under charge conjugation \cite{smirnov,etd} we find
\bea
\Im m \chi^{xx}(\omega,q)&\propto&
2\pi\sum_{n=1}^{[1/\xi]}Z_{2n}\ \delta(s^2 - M_{2n}^2) \nn
&& + \Re\re \frac{|F^{cos}[\theta(M_1,M_1,s)]_{11}|^2}{s\sqrt{s^2 -
4M_1^2}}\nn
&& + 2\Re\re
\frac{|F^{cos}[\theta(M,M,s)]_{+-}|^2}{s\sqrt{s^2-4M^2}}\nn
&& + \ldots
\label{cos}
\eea
where 
\be
\theta(m_1,m_2,s) = {\rm
arccosh}\left(\frac{s^2-m_1^2-m_2^2}{2m_1m_2}\right)\ .
\ee
Here the dependence of $\xi$ on the applied field is given by
\r{xi} and Fig.~\ref{fig:breathermass}. The first terms in \r{cos}
correspond to single-particle breather states. The (squares of the)
breather formfactors are given by 
\bea
Z_2&=& \frac{2(\sin 2\pi\xi)^2}{\cot\pi\xi}\ \nn
\times&&\exp\left[-2\int_0^\infty\frac{dx}{x}\ 
\frac{(\sinh{2\xi x})^2 \sinh x(1-\xi)}{\cosh x\ \sinh 2x\ \sinh\xi
x}\right] ,\nn
Z_4&=& \frac{2(\sin 4\pi\xi)^2}{(\cot\pi\xi)^2(\cot 3\pi\xi/2)^2}\
\nn 
\times&&\exp\left[-2\int_0^\infty\frac{dx}{x}\ 
\frac{(\sinh{4\xi x})^2 \sinh x(1-\xi)}{\cosh x\ \sinh 2x\ \sinh\xi
x}\right].
\eea
The soliton-antisoliton formfactor is
\bea
&&|F^{cos}_{+-}(\theta)|^2=\frac{(2\cot\pi\xi/2\
\sinh\theta)^2}{\xi^2} \frac{\cosh\theta/\xi+\cos\pi/\xi}
{\cosh 2\theta/\xi-\cos 2\pi/\xi}\nn
&&\times
\exp\left[-\int_0^\infty\frac{dx}{x}
\frac{[\cosh 2x \cos2x\theta/\pi -1] \sinh x(1-\xi)}{\cosh x\ \sinh 2x\
\sinh\xi x}\right],\nn
\eea
and the $B_1B_1$ breather-breather formfactor is
\bea
&&|F^{cos}_{11}(\theta)|^2=\lambda^4
\frac{(\sinh\theta)^2}{(\sinh\theta)^2+ (\sin\pi\xi)^2}\nn
&&\times
\exp\left[\!-8\!\int_0^\infty\!\frac{dx}{x}
\frac{\cos2x\theta/\pi \sinh\xi x \sinh x \sinh x(1+\xi)}{
(\sinh 2x)^2}\right]\!,\nn
\eea
where
\be
\lambda=2\cos\pi\xi/2\sqrt{2\sin\pi\xi/2}\ 
\exp\left[-\int_0^{\pi\xi}\frac{dx}{2\pi}\frac{x}{\sin x}\right].
\ee
The next most important contribution ({\sl i.e.} the term with the
lowest threshold) in the expansion \r{cos} comes from $B_1B_3$
breather-breather states. It will contribute at energies larger than
$M_1+M_3$, where $M_{1,3}$ are given by \r{mass}. 

In Fig.~\ref{fig:chixx35} we plot our result for $\chi^{xx}(\omega,q)$
as a function of $s/M=\sqrt{\omega^2-(\pi-q)^2}/M$ (where $M$ is given
by \r{solitonmass}) for an applied field of $H=3.5$ Tesla, which is
one of the values studied experimentally in \cite{dender}.

\begin{figure}[ht]
\begin{center}
\noindent
\epsfxsize=0.45\textwidth
\epsfbox{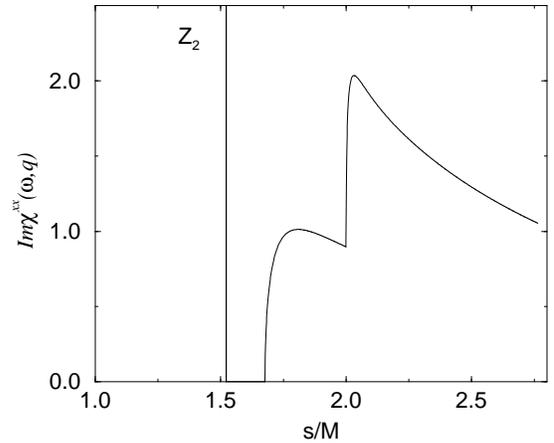}
\end{center}
\caption{\label{fig:chixx35}%
Transverse susceptibility $\chi^{xx}(s)$ for $H=3.5 {\rm Tesla}$.} 
\end{figure}

We see that there is one coherent mode at $s\approx 1.5 M$ with weight
$Z_2\approx 1.02$. At $s\approx 1.68 M$ the $B_1B_1$ continuum
appears and at $s=2M$ the soliton-antisoliton continuum. The
susceptibility is regular at both thresholds. 

Fig.~\ref{fig:chixx7} shows $\chi^{xx}(\omega,q)$ as for $H=7$
Tesla. The picture is qualitatively similar to the one at $3.5$ Tesla
although the soliton-antisoliton threshold has become more
pronounced. The weight of the delta-function corresponding to the
breather $B_2$ has diminished to $Z_2^\prime \approx 0.97$. 
As the field increases a singularity develops at the $s=2M$ threshold
until at $H\approx9.6$ Tesla the breather $B_4$ splits off the continuum. 
The situation at $H=14$ Tesla is shown in Fig.~\ref{fig:chixx14}. The
weight of $B_2$ is $Z_2^{\prime\prime}\approx 0.85$ whereas the fourth
breather has only a very small spectral weight of $Z_4\approx 0.01$.

\begin{figure}[ht]
\begin{center}
\noindent
\epsfxsize=0.45\textwidth
\epsfbox{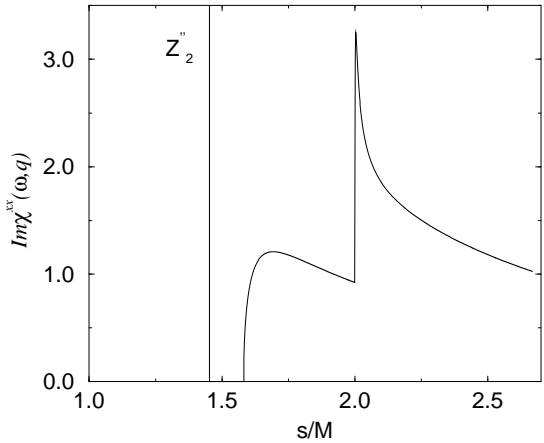}
\end{center}
\caption{\label{fig:chixx7}%
Transverse susceptibility $\chi^{xx}(s)$ for $H=7 {\rm Tesla}$.} 
\end{figure}

\begin{figure}[ht]
\begin{center}
\noindent
\epsfxsize=0.45\textwidth
\epsfbox{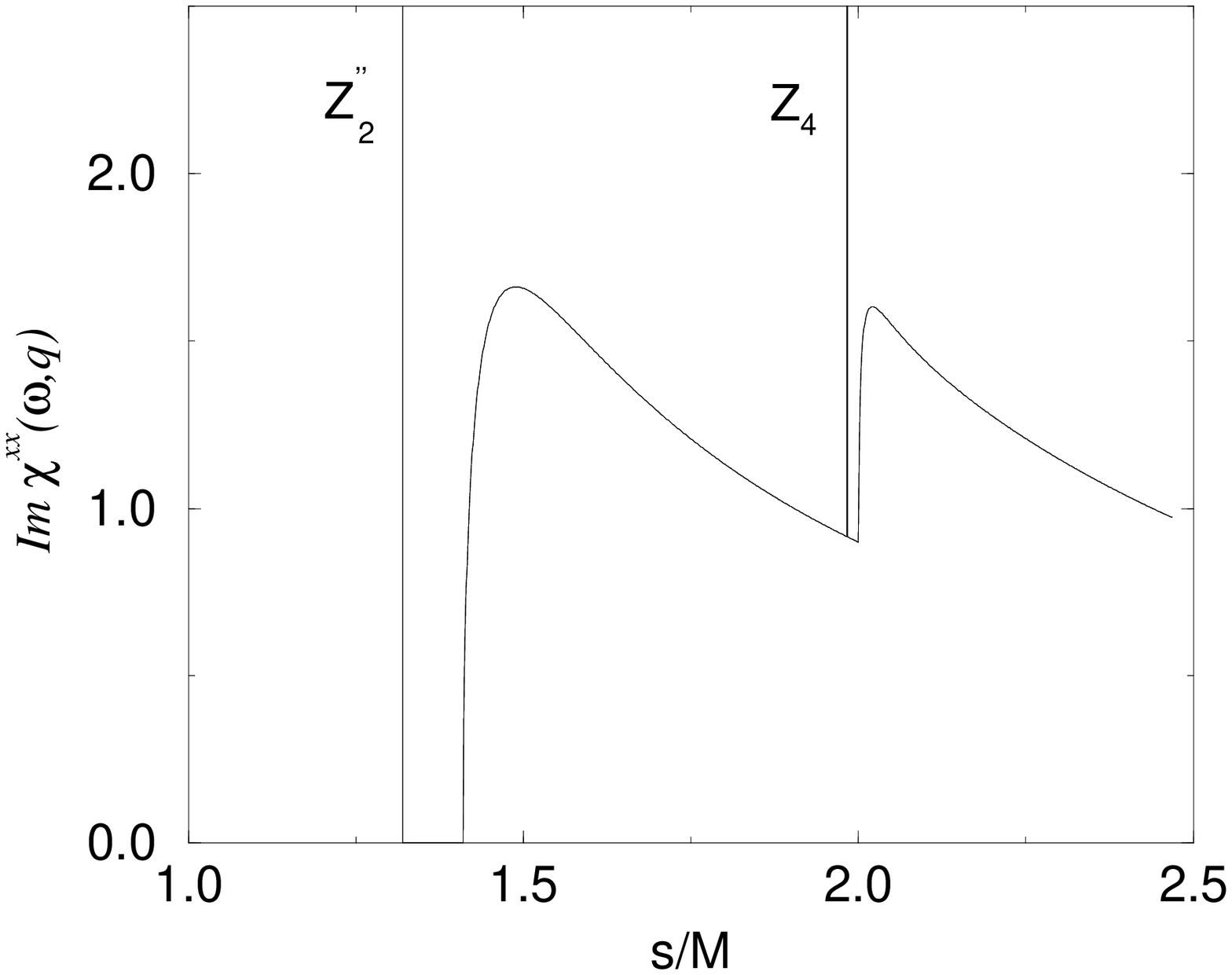}
\end{center}
\caption{\label{fig:chixx14}%
Transverse susceptibility $\chi^{xx}(s)$ for $H=14 {\rm Tesla}$.} 
\end{figure}

Let us now turn to $\Im m\chi^{yy}(\omega,q)$. It is clear from the
form of the Hamiltonian \r{hamil} that this will be different from
$\Im m\chi^{xx}(\omega,q)$. The formfactor expansion yields

\bea
\Im m\chi^{yy}(\omega,q)&\propto&
2\pi\sum_{n=1}^{[1/\xi]}Z_{2n-1}\ \delta(s^2 - M_{2n-1}^2)\nn
&& + 2\Re\re \frac{|F^{sin}[\theta(M,M,s)]_{+-}|^2}{s\sqrt{s^2 -
4M^2}}\nn
&&+ 2\Re\re \frac{|F^{sin}[\theta(M_1,M_2,s)]_{12}|^2}
{\sqrt{(s^2-M_1^2-M_2^2)^2-4M_1^2M_2^2}}\nn
&& + \ldots
\label{sin}
\eea
Here the breather formfactors are given by
\bea
Z_1&=& \frac{8(\cos \pi\xi/2)^4}{\cot\pi\xi/2}\ \nn
\times&&\exp\left[-2\int_0^\infty\frac{dx}{x}\ 
\frac{\sinh\xi x\ \sinh x(1-\xi)}{\cosh x\ \sinh 2x}\right] ,\nn
Z_3&=& \frac{4\sin 3\pi\xi\ (\sin 3\pi\xi/2)^2}{(\cot\pi\xi)^2}\
\nn
\times&&\exp\left[-2\int_0^\infty\frac{dx}{x}\ 
\frac{(\sinh{3\xi x})^2 \sinh x(1-\xi)}{\cosh x\ \sinh 2x\ \sinh\xi
x}\right] .
\eea
The soliton-antisoliton formfactor is found to be
\bea
&&|F^{sin}_{+-}(\theta)|^2=\frac{(2\cot\pi\xi/2\
\sinh\theta)^2}{\xi^2} \frac{\cosh\theta/\xi-\cos\pi/\xi}
{\cosh 2\theta/\xi-\cos 2\pi/\xi}\nn
&&\times
\exp\left[-\int_0^\infty\frac{dx}{x}
\frac{[\cosh 2x \cos2x\theta/\pi -1] \sinh x(1-\xi)}{\cosh x\ \sinh 2x\
\sinh\xi x}\right].\nn
\eea
Finally we take into account the $B_1B_2$ breather-breather state
which has a formfactor of
\bea
&&|F^{sin}_{12}(\theta)|^2=\frac{\tan\pi\xi}{2}
|g(\theta-i\pi\xi/2)g(\theta+i\pi\xi/2)|^2\lambda^6\nn
&&\times\left(1+\frac{1}{4\cos
\pi\xi/2(\cosh\theta+\cos\pi\xi/2)}\right)^2\nn
&&\times\exp\left[\!-8\!\int_0^\infty\!\!\frac{dx}{x}
\frac{\cos2x\theta/\pi \sinh 2\xi x \sinh x \sinh x(1+\xi)}{
(\sinh 2x)^2}\right]\nn
&&\times
\exp\left[\!-8\!\int_0^\infty\!\frac{dx}{x}
\frac{\cosh 2x\xi \sinh\xi x \sinh x \sinh x(1+\xi)}{
(\sinh 2x)^2}\right],\nn
\eea
where
\be
g(x)= \frac{\sinh x}{\sinh x - i\sin\pi\xi}\ .
\ee
The next most important contribution to \r{sin} is due to $B_1B_1B_1$
three breather states with a threshold at $3M_1$.

In Figs.~\ref{fig:chiyy35}-\ref{fig:chiyy14} we plot $\chi^{yy}(s)$
for three different values of the applied field, namely $H=3.5,\ 7,\
14$ Tesla. The emerging picture remains qualitatively unchanged for
all three values of $H$. There are two coherent modes corresponding to
the breathers $B_1$ and $B_3$. At $s=2M$ the soliton-antisoliton
continuum starts and at $s=M_1+M_2$ we observe the onset of the
$B_1B_2$ breather-breather continuum. The latter one exhibits a 
singularity as a function of $s$. The weight of the delta
functions corresponding to the breather $B_1$ decrease with increasing 
field: $Z_1\approx 2.07$, $Z_1^\prime\approx 2.04$ and
$Z_1^{\prime\prime}=1.95$. For $B_3$ we find a decrease with
increasing field after an initial increase: $Z_3\approx 0.34$,
$Z_3^\prime\approx 0.37$, $Z_3^{\prime\prime}=0.36$. Note that the
spectral weight of the heavier $B_3$ particle is always significantly
smaller.

\begin{figure}[ht]
\begin{center}
\noindent
\epsfxsize=0.45\textwidth
\epsfbox{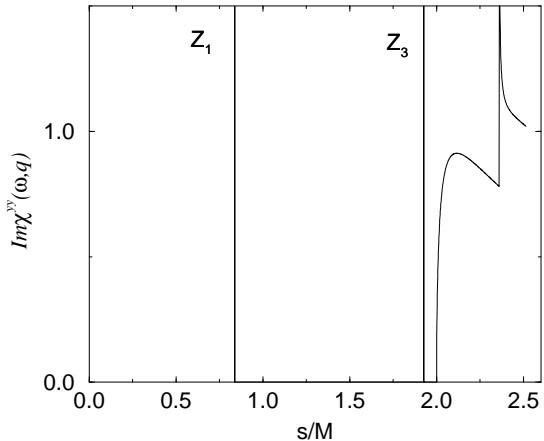}
\end{center}
\caption{\label{fig:chiyy35}%
Transverse susceptibility $\chi^{yy}(s)$ for $H=3.5 {\rm Tesla}$.} 
\end{figure}

\begin{figure}[ht]
\begin{center}
\noindent
\epsfxsize=0.45\textwidth
\epsfbox{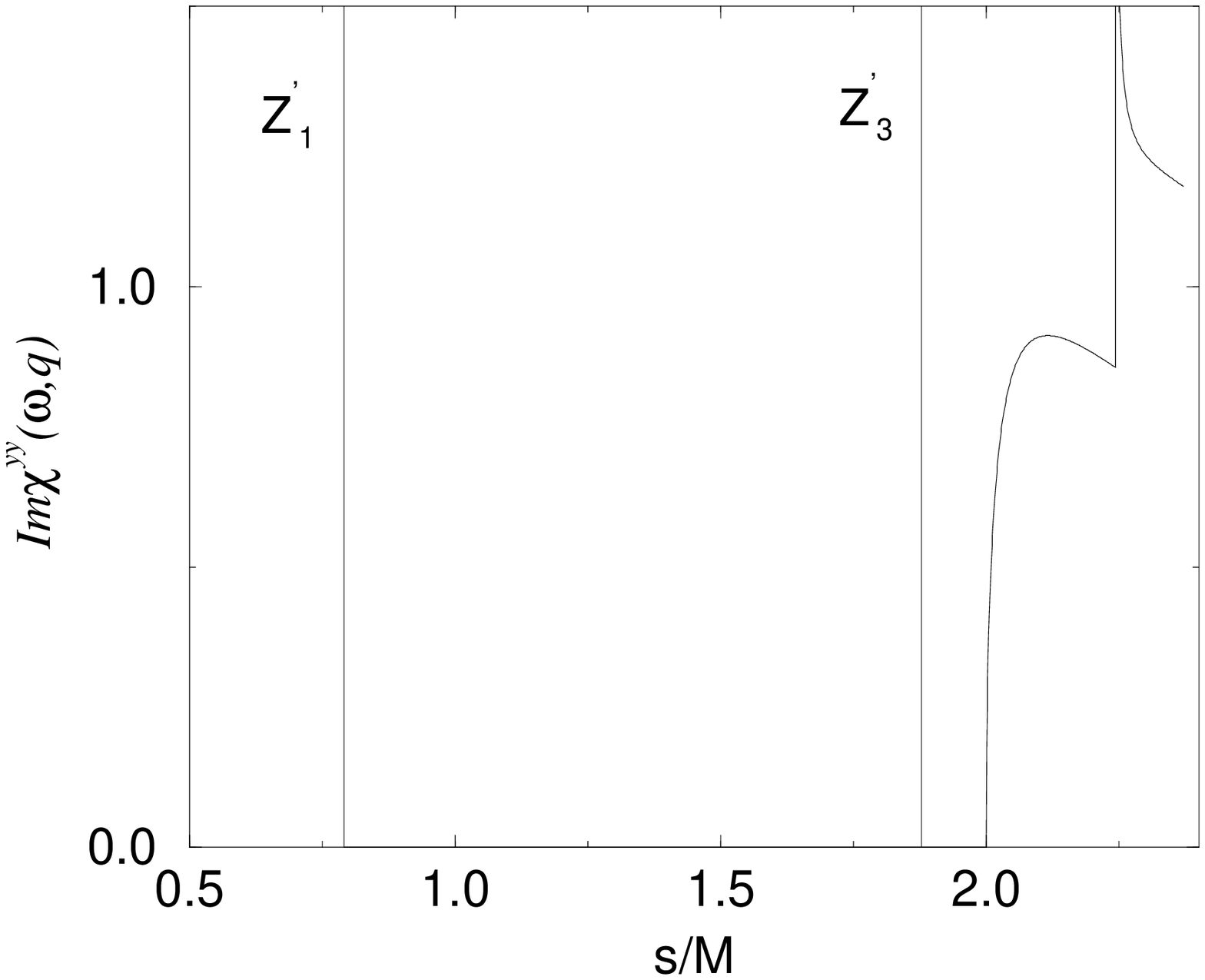}
\end{center}
\caption{\label{fig:chiyy7}%
Transverse susceptibility $\chi^{yy}(s)$ for $H=7 {\rm Tesla}$.} 
\end{figure}

\begin{figure}[ht]
\begin{center}
\noindent
\epsfxsize=0.45\textwidth
\epsfbox{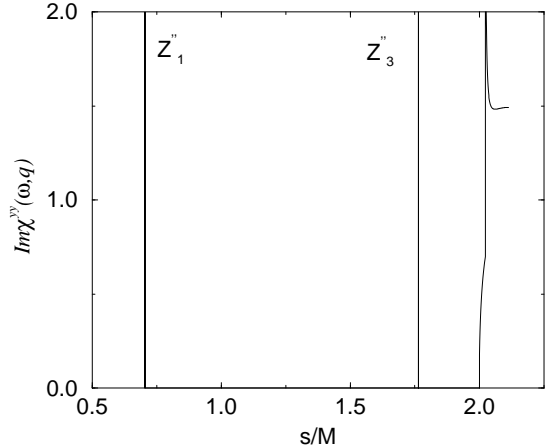}
\end{center}
\caption{\label{fig:chiyy14}%
Transverse susceptibility $\chi^{yy}(s)$ for $H=14 {\rm Tesla}$.} 
\end{figure}
Let us now compare our results with the experimental findings. Our
results for the imaginary part of the dynamical susceptibilities are
consistent with the experiment at $7$ Tesla (Fig. 3 (c) of
\cite{dender}): as the beam was unpolarized the experiment observed
$\frac{1}{2}(\chi^{xx}+\chi^{yy})$ (the contribution from $\chi^{zz}$
emerges only at energies greater than $g \mu_B H$).
The observed peaks at $0.17 {\rm meV}$, $0.34 {\rm meV}$ and $0.44
{\rm meV}$ correspond to the three breathers $B_1$, $B_2$ and $B_3$
and agree very well with our prediction for the mass spectrum \r{mass}
and \r{solitonmass} (the first two peaks were already discussed in
\cite{oa}). In order to compare our calculated weights of the
delta-functions in the susceptibilities to the experimentally observed
intensities one needs to convolve with the instrumental resolution. 
Convolving our exact results with a Gaussian we obtain a fit to the
experimental data shown in Fig.~\ref{fig:fit}.
\begin{figure}[ht]
\begin{center}
\noindent
\epsfxsize=0.45\textwidth
\epsfbox{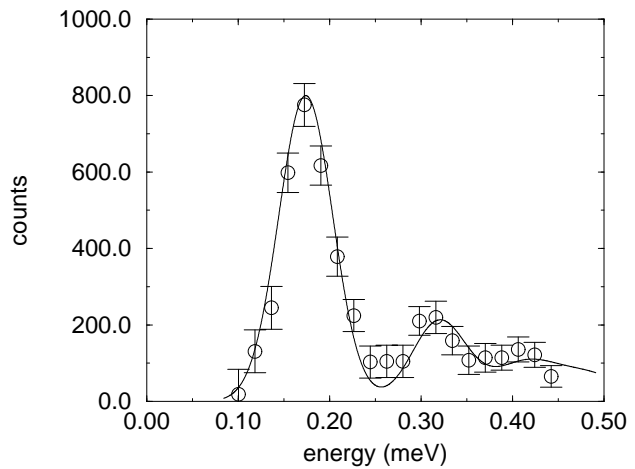}
\end{center}
\caption{\label{fig:fit}%
Calculated (solid line) vs measured (data points taken from
Ref. [2]) intensity at $H=7 {\rm Tesla}$.}  
\end{figure}
We see that the theory is in good agreement with experiment.

\begin{center}
{\bf Acknowledgements}
\end{center}

We are grateful to G. Aeppli, C. Broholm, R. Coldea, R. Cowley,
A. Millis and L.P. Regnault for important discussions. F.H.L.E. was
supported by the EU under Human Capital and Mobility fellowship grant
ERBCHBGCT940709.

\end{narrowtext}


\begin{thebibliography}{99}

\bibitem{date}
M. Date {\sl et.al.}, Suppl. Prog. Theor. Phys. {\bf 46} (1970) 194.

\bibitem{dender}
D. C. Dender, P.R. Hammar, D.H. Reich, C. Broholm and G. Aeppli,
Phys. Rev. Lett. {\bf 79} (1997) 1750.

\bibitem{oa2}
M. Oshikawa, M. Yamanaka and I. Affleck, \PRL{78}{1997}{1984}.

\bibitem{oa}
M. Oshikawa and I. Affleck, preprint {\tt cond-mat/9706085}.

\bibitem{vladb}
V.~E. Korepin, A.~G. Izergin, and N.~M. Bogoliubov, {\em {Quantum Inverse
  Scattering Method, Correlation Functions and Algebraic Bethe Ansatz}}
  (Cambridge University Press, 1993).

\bibitem{spectrum}
V.E. Korepin and L.D. Faddeev, Theor. Mat. Phys. {\bf 25} (1975) 1039,\\
R. Dashen, B. Hasslacher and A. Neveu, Phys. Rev. {\bf D11} (1975)
3424.

\bibitem{zamo2}
A.B. Zamolodchikov and Al.B. Zamolodchikov, Annals of Physics {\bf
120} (1979) 253.

\bibitem{mtm}
H. Bergknoff and H. Thacker, \PRD{19}{1979}{3666},\\
V.E. Korepin, \TMP{41}{1979}{169}.

\bibitem{luth}
A. Luther, \PRB{14}{1976}{2153}.

\bibitem{aff}
I. Affleck,  in {\sl Fields, Strings and Critical Phenomena}, {\sl Les
Houches, Session XLIX}, edited by E. Brezin and J. Zinn-Justin
(North-Holland, Amsterdam, 1988).

\bibitem{book}
A.M. Tsvelik, {\sl Quantum Field Theory in Condensed Matter Physics},
Cambridge University Press (1995).

\bibitem{nst}
I. Affleck, \NPB{265}{1986}{409},\\
F.D.M. Haldane, \PRB{36}{1987}{5291},\\
D.G. Shelton, A.A. Nersesyan, A.M. Tsvelik, \PRB{53}{1996}{8521}.

\bibitem{karowski} 
M. Karowski and P. Weisz, Nucl. Phys. {\bf B139} (1978) 455,\\
B. Berg, M. Karowski, P. Weisz, \PRD{19}{1979}{2477}.

\bibitem{smirnov}
F.~A. Smirnov, {\sl Form Factors in Completely Integrable Models of
Quantum Field Theory},\hfill\break World Scientific (1992).

\bibitem{lukyanov}
S. Lukyanov, Comm. Math. Phys. {\bf 167} (1995) {183},\\
S. Lukyanov and A.B. Zamolodchikov, \NPB{493}{1997}{2541},\\
S. Lukyanov, preprint {\tt hep-th/9703190}.

\bibitem{etd}
F.H.L. E\char'31ler, A.M. Tsvelik and G. Delfino, \PRB{56}{1997}{11001}.

\end{thebibliography}
\end{document}